\documentclass{optica-article}
\journal{opticajournal}
\articletype{Research Article}

\usepackage{lineno}
\usepackage{graphicx}
\usepackage{setspace}
\usepackage{tocloft}
\usepackage{subcaption}

\begin{document}

\title{Experimental Validation of Double-Blind Fourier Holography in an HWO-Like Segmented Aperture}

\author{Oren Pedatzur,\authormark{1,*} Noah Schwartz,\authormark{2,†}, D. Scott Acton\authormark{2,‡} and Ben Buky\authormark{2}}

\address{\authormark{1}Villanova University, Department of Physics, 800 E. Lancaster Ave, Villanova, 19085, PA, USA\\
\authormark{2}UK Astronomy Technology Centre, Edinburgh EH9 3HJ, United Kingdom}

\email{\authormark{*}orenpedatzur@gmail.com}
\email{\authormark{†}noah.schwartz@stfc.ac.uk}
\email{\authormark{‡}scottjwst@gmail.com}

\begin{abstract*}
Double-Blind Fourier Holography (DBFH) is a phase-retrieval method that has been shown numerically to address wavefront-sensing challenges for future segmented-aperture space telescopes, including the Habitable Worlds Observatory (HWO). DBFH uses a set of carefully selected pupil-plane amplitude masks and their corresponding focal-plane intensity measurements to construct an algebraically linear inverse problem, enabling computationally efficient and stable reconstruction of the complex pupil field. Here, we present the first experimental demonstration of DBFH using a spatial light modulator and an HWO EAC5-like segmented-aperture geometry. We reconstruct challenging pseudo-random wavefronts and compare DBFH with the established Focus-Diverse Phase Retrieval (FDPR) method. The two methods achieve comparable reconstruction precision, while exhibiting distinct strengths and limitations. We further demonstrate DBFH wavefront sensing in the presence of a binary source, a scenario that is not naturally accommodated by conventional FDPR. These results provide experimental proof of concept and establish technology readiness level 3 for DBFH as a wavefront-sensing method for space telescopes with segmented primary mirrors.
\end{abstract*}

\section{Introduction} \label{sect:intro}

Space telescopes provide unparalleled opportunities to address some of the deepest unanswered questions in astronomy: How did the Universe begin? What drives cosmic expansion? Are there habitable planets beyond our own, and does life exist elsewhere in the Universe? We are fortunate to be witnessing a remarkable era in astronomy, in which successive generations of flagship space telescopes continue to deliver new capabilities and insights, bringing us closer to answering these fundamental questions. The recent launch of the Nancy Grace Roman Space Telescope marks the latest chapter in this progression, while attention is already turning toward the next generation of flagship observatories. Looking forward, the Habitable Worlds Observatory (HWO), planned for launch in the 2040s, will have as one of its primary goals the direct imaging of potentially habitable, Earth-like exoplanets and the characterization of their atmospheres for possible signs of life \cite{NASEM2021Astro2020}. Building on strong heritage from previous missions like James Webb (JWST) and Roman space telescopes, current HWO concepts combine a large, segmented primary mirror with high-performance coronagraphy instrument to suppress light from a host star and enable detection of extremely faint nearby planets. Achieving the required coronagraphic contrast places stringent demands on the stability of the telescope optical system, including picometer-scale wavefront stability over relevant observation timescales \cite{juanola2019evaluating,juanola2022modeling}. Meeting this requirement will require both highly stable structural and thermal designs and sensitive wavefront-sensing and control techniques.

Focal-plane wavefront sensing is particularly attractive for space telescopes because it measures aberrations through the same optical path used to form the science image, thereby reducing reliance on separate metrology systems. Focus-Diverse Phase Retrieval (FDPR) \cite{gonsalves1982phase,gonsalves2018phase} is one of the most established approaches to focal-plane phase retrieval. In FDPR, intensity measurements acquired with known amounts of defocus, are used to estimate the pupil-plane wavefront. FDPR techniques have played an important role in space-telescope wavefront sensing and are currently used for fine phasing of JWST, where nanometer-scale residual wavefront errors have been demonstrated \cite{acton2004james,acton2022phasing}. FDPR is typically formulated as a nonlinear inverse problem and solved through iterative numerical optimization over an aberration model, often expressed using a truncated Zernike basis, and can incur substantial computational cost.

Double-Blind Fourier Holography (DBFH) provides an alternative approach to phase retrieval. DBFH was first introduced by Raz et al. for one-dimensional signals \cite{raz2011vectorial,raz2013vectorial} and subsequently extended to lensless imaging \cite{raz2014direct}. Variants of the method have since been applied to problems including coherent diffraction imaging of nanoparticles using X-ray free-electron lasers \cite{leshem2016direct} and attosecond pulse characterization \cite{pedatzur2019double}. More recently, DBFH was proposed as a focal-plane wavefront-sensing method for segmented space telescopes and was numerically investigated under conditions relevant to HWO \cite{pedatzur2026double}.

In the wavefront-sensing implementation of DBFH, the pupil is divided into complementary regions using known amplitude masks, and focal-plane intensity measurements are acquired for different pupil configurations. Linear combinations of these measurements provide the interference cross term between the fields originating from the complementary sub-apertures. A compact-support constraint encodes the known finite spatial extent of each sub-aperture, requiring its associated pupil-plane field to vanish everywhere outside that region. The measured cross term, together with the known compact-support constraints imposed by the pupil masks, forms an overdetermined system of linear equations whose solution yields the complex focal-plane fields. A straightforward inverse Fourier transform then recovers the complex pupil field and hence the wavefront. DBFH retrieves the wavefront phase \textit{in situ} from in-focus intensity measurements using a non-iterative linear reconstruction, without requiring the deliberate defocus used in FDPR or restricting the wavefront to a low-dimensional aberration basis.

In this work, we present the first experimental validation of DBFH on an HWO-like segmented aperture. Using an optical testbed based on a programmable spatial light modulator (SLM), we generate a 37-segment aperture based on the HWO EAC5 geometry and impose controlled, nontrivial wavefront errors across the pupil. We reconstruct these wavefronts using DBFH and compare the results directly with FDPR under similar experimental conditions. We further investigate the dependence of DBFH reconstruction error on signal-to-noise ratio (SNR) and compare the measurements with numerical predictions. Finally, we demonstrate DBFH wavefront sensing in a field containing two mutually incoherent point sources. Because DBFH uses in-focus intensity measurements, the two PSFs can remain spatially separated on the detector and their wavefront estimates can be independently recovered and combined.

This manuscript is organized as follows. Section~\ref{section: DBFH} reviews the mathematical formulation, assumptions, and implementation of DBFH for wavefront sensing. Section~\ref{section: experimetnal setup} describes the experimental setup and data-acquisition procedure. Section~\ref{section: results} presents the experimental DBFH reconstructions, comparison with FDPR, and dependence on signal-to-noise ratio. Section~\ref{section: binary source} demonstrates wavefront sensing using a binary source, and Section~\ref{section: conclusions} summarizes the results and discusses their significance. 

\section{DBFH for Wavefront Sensing} \label{section: DBFH}
Double Blind Fourier Holography is a phase retrieval method which uses diffraction images originating from two mutually coherent sources associated with regions of finite extent. When applied to the problem of wavefront sensing (WFS), the field we wish to characterize is the wavefront at the telescope pupil plane, and the diffraction images correspond to intensity in the focal plane. By exposing complementary regions of the pupil we effectively separate the complex field at the pupil plane into two components:
\begin{equation}
    E(\mathbf{x}) = E_1(\mathbf{x}) + E_2(\mathbf{x}),
\end{equation}
with $E_{i}(\mathbf{x})$ vanishing outside their corresponding complementary (non-overlapping) areas of compact support (CS, or sub-apertures) $CS_{i}$, such that $CS_1 \cup CS_2 = pupil$. We use $\mathbf{x}$ for coordinates in the pupil plane. The fields are propagated to the focal plane a using a 2D Fourier transform:
\begin{gather}
    A_1(\boldsymbol{\xi}) = \mathcal{F}[E_1(\mathbf{x})],\\
    A_2(\boldsymbol{\xi}) = \mathcal{F}[E_2(\mathbf{x})],\\
    A(\boldsymbol{\xi}) = \mathcal{F}[E(\mathbf{x})] = A_1(\boldsymbol{\xi}) + A_2(\boldsymbol{\xi}).
\end{gather}
where we use $A_i$ for the focal plane fields and $\boldsymbol{\xi}$ for focal plane coordinates. A focal plane intensity image from the fully exposed pupil is simply a coherent sum of $A_1$ and $A_2$:
\begin{gather}
    I_1(\boldsymbol{\xi}) = |A_1(\boldsymbol{\xi})|^2, \label{eq: intensity input 1} \\     
    I_2(\boldsymbol{\xi}) = |A_2(\boldsymbol{\xi})|^2, \label{eq: intensity input 2} \\
    I(\boldsymbol{\xi}) = |A_1(\boldsymbol{\xi})+A_2(\boldsymbol{\xi})|^2 \label{eq: intensity input 3}
\end{gather}
where $I_i$ denote the intensity corresponding to the sub-apertures. The goal in this phase retrieval problem is to estimate the focal plane phases:
\begin{gather}
    \phi_1(\boldsymbol{\xi}) = \arg(A_1(\boldsymbol{\xi})), \\
    \phi_2(\boldsymbol{\xi}) = \arg(A_2(\boldsymbol{\xi})).
\end{gather}
We will then inverse Fourier transform the focal plane field to get the pupil plane field and hence the wavefront. 

For a straight boundary between $CS_1$ and $CS_2$, these three intensity images, eqs. (\ref{eq: intensity input 1}-\ref{eq: intensity input 3}), together with the known supports are sufficient inputs for DBFH as was demonstrated in \cite{pedatzur2026differential}. For computational convenience we add a fourth in-quadrature measurement in which we apply a quarter-wavelength retardation over one sub-aperture:
\begin{equation}
    I_{\pi/2}(\boldsymbol{\xi}) = |A_1(\boldsymbol{\xi})+iA_2(\boldsymbol{\xi})|^2
\end{equation}
Applying this $\pi/2$ phase can be easily achieved in adaptive optics systems and specifically segmented mirror telescopes. The interference images allow us to extract linear combinations of cross-terms involving $A_1$ and $A_2$:
\begin{gather}
    C_0(\mathbf{k}) = I - I_1 - I_2 = A_1 A_2^* + A_1^*A_2 \\
    C_{\pi/2}(\mathbf{k}) = I_{\pi/2} - I_1 - I_2 = iA_1^*A_2 - iA_1 A_2^*
\end{gather}
A single cross-term is easily isolated like so:
\begin{equation}
    S = \tfrac{1}{2}(C_0 + iC_{\pi/2}) = A_1 A_2^*
\end{equation}
As mentioned above, a straight boundary between $CS_1$ and $CS_2$ enables the recovery of $A_1 A_2^*$ from $Re(A_1 A_2^*)$ using an algorithmic step akin to the Kramers-Kronig relations.

DBFH formulates a linear system of equations whose unknowns are the focal plane phasors
$X_1(\boldsymbol{\xi}) = e^{i\phi_1(\boldsymbol{\xi})}$ and
$X_2(\boldsymbol{\xi}) = e^{i\phi_2(\boldsymbol{\xi})}$
for each spatial frequency. One subset of the linear equations enforces consistency with the measured cross-term:
\begin{equation}\label{eq:subset-cross}
    S X_2 - |A_1||A_2| X_1 = 0.
\end{equation}
Two additional subsets of linear equations enforce the sub-apertures' CS constraints. Specifically, the inverse Fourier transform of each focal plane field must vanish outside the corresponding CS:
\begin{gather}
    E_1(\mathbf{x}) = \mathcal{F}^{-1}\left[|A_1(\boldsymbol{\xi})| X_1(\boldsymbol{\xi})\right](\mathbf{x}) = 0
    \quad \text{for} \quad \mathbf{x} \notin CS_1 \label{eq:CSA}\\
    E_2(\mathbf{x}) = \mathcal{F}^{-1}\left[|A_2(\boldsymbol{\xi})| X_2(\boldsymbol{\xi})\right](\mathbf{x}) = 0
    \quad \text{for} \quad \mathbf{x} \notin CS_2 \label{eq:CSB}
\end{gather}
Adequate sampling in the focal plane is required to ensure that these three subsets of equations together form an overdetermined linear system.

This homogeneous system is invariant under multiplication of both unknown fields by an arbitrary nonzero complex scalar. We remove this common scale and global-phase ambiguity by fixing one component of $X_1(\boldsymbol{\xi}) = 1$. Specifically, we select a well-measured focal-plane pixel to serve as an anchor, and impose the gauge condition $X_1(\boldsymbol{\xi_{anchor}}) = 1$. This converts the homogeneous system into the nonhomogeneous linear system which we solve in the least-squares sense, recovering estimates for both $X_1$ and $X_2$. Because the unit-modulus constraints on $X_1$ and $X_2$ are relaxed during the linear solve, the recovered phasor solutions are finally projected onto the unit circle: 
\begin{equation}
    \hat X_i(\boldsymbol{\xi}) =  \frac{X_i(\boldsymbol{\xi})}{|X_i(\boldsymbol{\xi})|}
\end{equation}

\section{Experimental Setup and Data Acquisition} \label{section: experimetnal setup}

\begin{figure}
    \centering
    \includegraphics[width=0.9\linewidth]{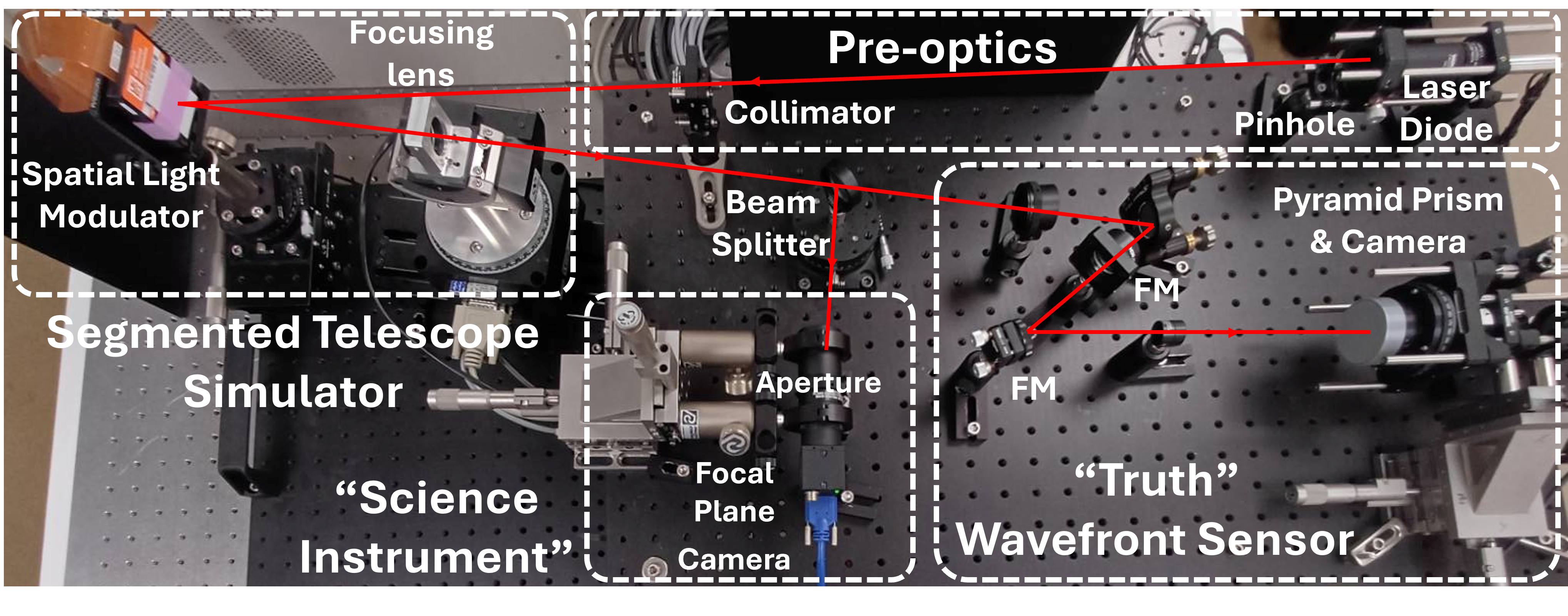}
    \vspace{1mm}
    \caption{\textbf{Experimental Setup.} The NESSIE (Next-generation Extreme AO and Segmented Systems for Instrumentation in Edinburgh) adaptive optics testbed at the UK Astronomy Technology Centre (UKATC). The testbed is designed for the investigation of phasing techniques for segmented space telescopes, with aberrations introduced by a spatial light modulator (SLM) and estimated from focal-plane images.}
    \label{fig:experimental setup}
\end{figure}

We used the experimental setup shown in Fig.~\ref{fig:experimental setup} to validate the DBFH methodology. The system is based on a spatial light modulator (SLM), which facilitates the implementation of both DBFH and FDPR on a pupil representative of the HWO EAC-5 architecture. This configuration enables a direct comparison of the two wavefront sensing techniques under identical experimental conditions.

The optical bench consists of a source module, a programmable phase modulator, and an imaging arm. The source module comprises a 635 nm laser diode (Thorlabs CPS635R) coupled to a 10~\textmu m pinhole, which acts as a spatial filter. The diverging beam propagates 500 mm before being collimated by an achromatic doublet (Thorlabs AC254-500A). The pupil shape and wavefront aberrations are introduced using a Hamamatsu X15213-01 liquid crystal-on-silicon (LCOS) spatial light modulator (SLM). The device provides programmable phase modulation through a two-dimensional array of addressable electrodes that locally modify the refractive index of a liquid-crystal layer. The SLM comprises a $1024 \times 1272$ pixel array with a 12.5~\textmu m pixel pitch, corresponding to a maximum active aperture of approximately 12.8 mm. The collimated beam is incident on the SLM at $5^\circ$ to the surface normal. A linear phase ramp is superimposed on all phase patterns, steering the first diffracted order by $1.45^\circ$. This separates the modulated beam from unwanted light arising from the finite pixel fill factor, specular reflection, and higher diffraction orders, allowing these components to be rejected with a spatial filter. Following the SLM, the beam is relayed by a second achromatic doublet (Thorlabs AC254-500A) and directed to the imaging arm via a 50:50 plate beamsplitter (Thorlabs BSW10R). The reflected beam is focused onto the detector by a plano-convex lens (Thorlabs LA1708). Images are recorded using a Point Grey Flea3 CMOS camera with a $1280 \times 1024$ pixel array and a pixel pitch of 5.3~\textmu m. For a 12 mm pupil diameter, the image sampling is approximately 4.08 pixels per $\lambda/D$. Pixel binning is subsequently applied to achieve near-Nyquist sampling. The optical bench also incorporates a truth-sensing arm, which was not directly used for the measurements reported in this work. Its purpose is to independently verify the wavefront generated by the SLM. The arm employs a $200 \times 200$ pyramid wavefront sensor, enabling high-resolution wavefront characterization.
 
\begin{figure}
    \centering
    \includegraphics[width=0.9\linewidth]{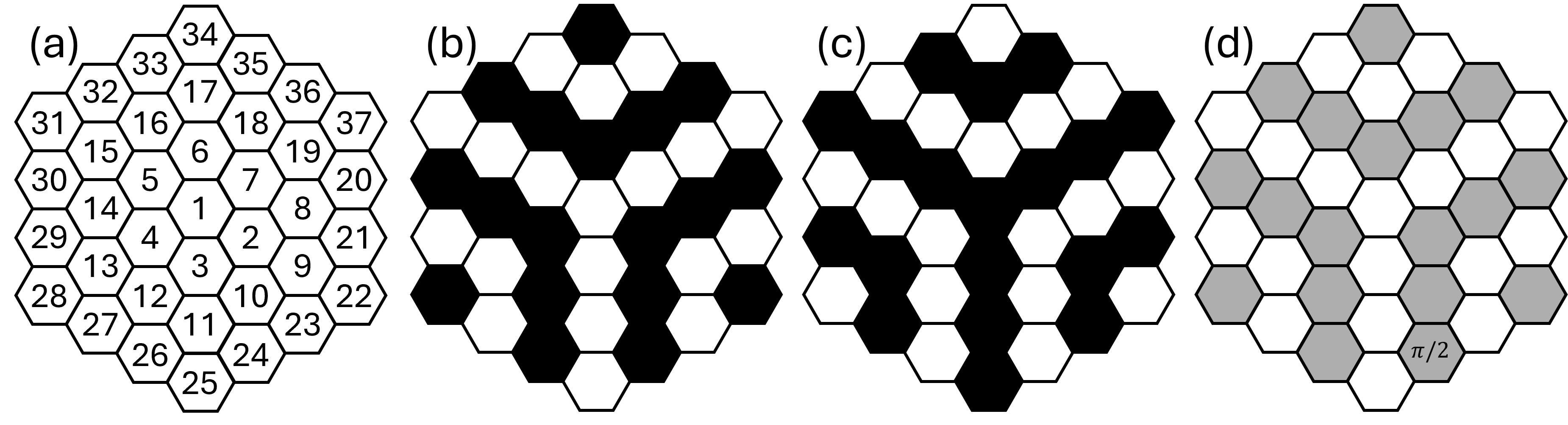}
    \vspace{1mm}
    \caption{\textbf{Aperture configurations used in the DBFH demonstration.}(a) Fully exposed 37-segment HWO EAC5 aperture with segment numbering. (b,c) Complementary amplitude masks exposing the odd- and even-numbered segments, respectively; exposed segments are shown in white and blocked segments in black. (d) Fully exposed aperture, with a $\pi/2$ phase shift applied to the even-numbered segments (gray), corresponding to the in-quadrature measurement.}
    \label{fig:masks}
\end{figure}

For consistency with previous results \cite{pedatzur2026double} we chose to implement DBFH using odd and even segment number sub-apertures, as in fig. \ref{fig:masks}. This strategy samples the pupil area with uniform density similarly in both odd and even sub-apertures. Other choices may be optimal in other aspects, e.g. left-right allows skipping the 4th in-quadrature capture as discussed in section \ref{section: DBFH}. Importantly, we must allow a thin occulting vane between neighboring hexagons to avoid spill-over of energy between the odd and even sub-apertures due to the size of the smallest phase ramp unit in the SLM (3x3 SLM pixels).

Each experiment comprises two pupil-plane phase states: (i) an initial reference state with zero applied phase and (ii) a nontrivial phase pattern to be recovered by wavefront sensing. For each phase state, we acquire the four measurements required for DBFH: odd segments, even segments, full-aperture, and in-quadrature captures, as well as the five FDPR images corresponding to $-8, -4, 0, +4$, and $+8$ waves defocus. All DBFH measurements must be acquired using the same exposure time or, equivalently, normalized by their respective exposure times. To vary the effective peak intensity and signal-to-noise ratio (SNR), the acquisition is repeated using 1, 10, 100, and 1000 co-added frames for every capture. Each wavefront-sensing method is then applied independently to the zero-phase and nontrivial-phase data sets. The recovered zero-phase estimate is subtracted from the recovered nontrivial phase, and the resulting differential phase is compared with the phase pattern applied to the SLM, which serves as the ground truth.

\section{Results} \label{section: results}

\begin{figure}
    \centering
    \includegraphics[width=0.9\linewidth]{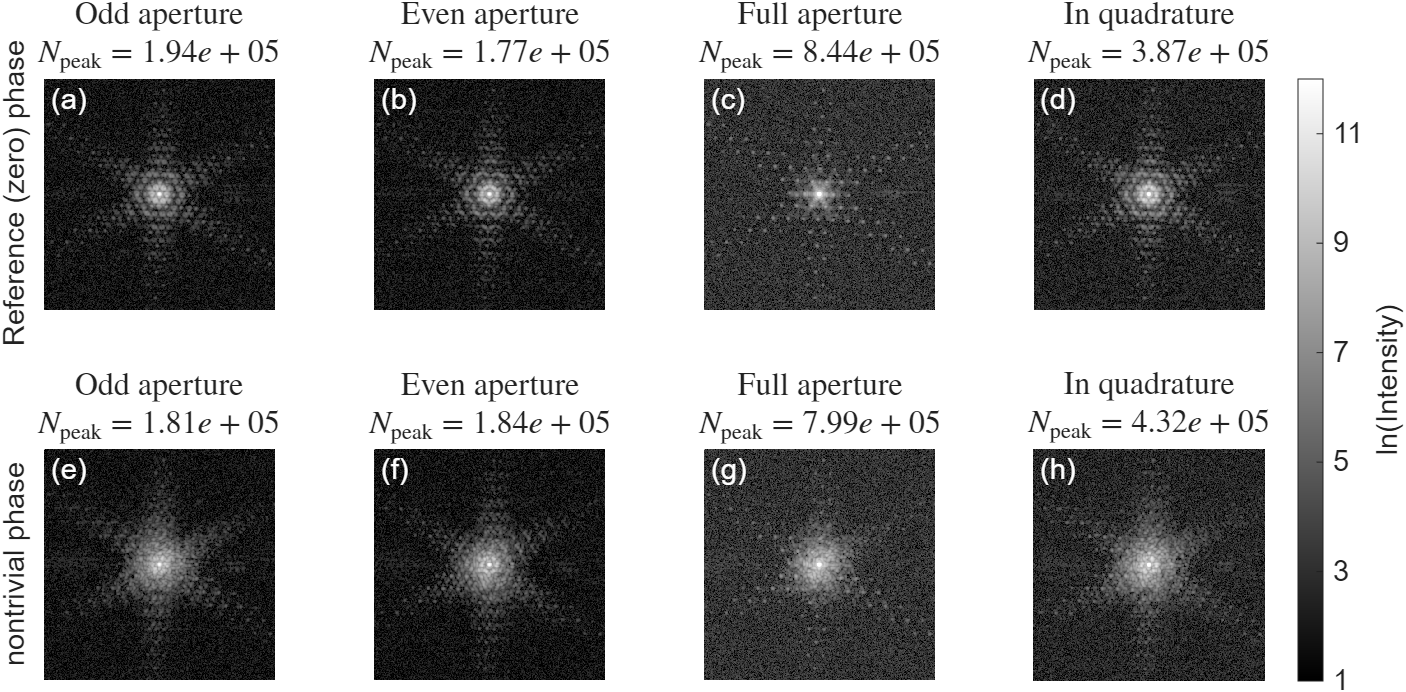}
    \vspace{1mm}
    \caption{\textbf{DBFH input intensity images.} Measured focal-plane intensities for the four aperture configurations shown in Fig.~2: odd-segment aperture, even-segment aperture, full aperture, and the in-quadrature aperture with a $\pi/2$ phase shift applied to the even-numbered segments. The top row (a--d) shows measurements with the reference zero SLM phase, and the bottom row (e--h) shows measurements with the non-trivial test phase. Intensities are displayed on a logarithmic scale, and the peak photon count, $N_{\mathrm{peak}}$, is reported above each image.}
    \label{fig:DBFH PSFs}
\end{figure}

We chose to demonstrate DBFH in a challenging scenario which includes up to 3rd order random Zernike components, and up to a single wavelength WFE. Figures \ref{fig:DBFH PSFs}(a-d) show the focal-plane intensity images, corresponding to the odd,even, full, and in-quadrature inputs to DBFH for the reference zero phase, with 1000 co-added frames. Similarly, figures \ref{fig:DBFH PSFs}(e-h) are the inputs with the applied nontrivial WFE, exhibiting far less symmetric diffraction pattern.

\begin{figure}
    \centering
    \includegraphics[width=1\linewidth]{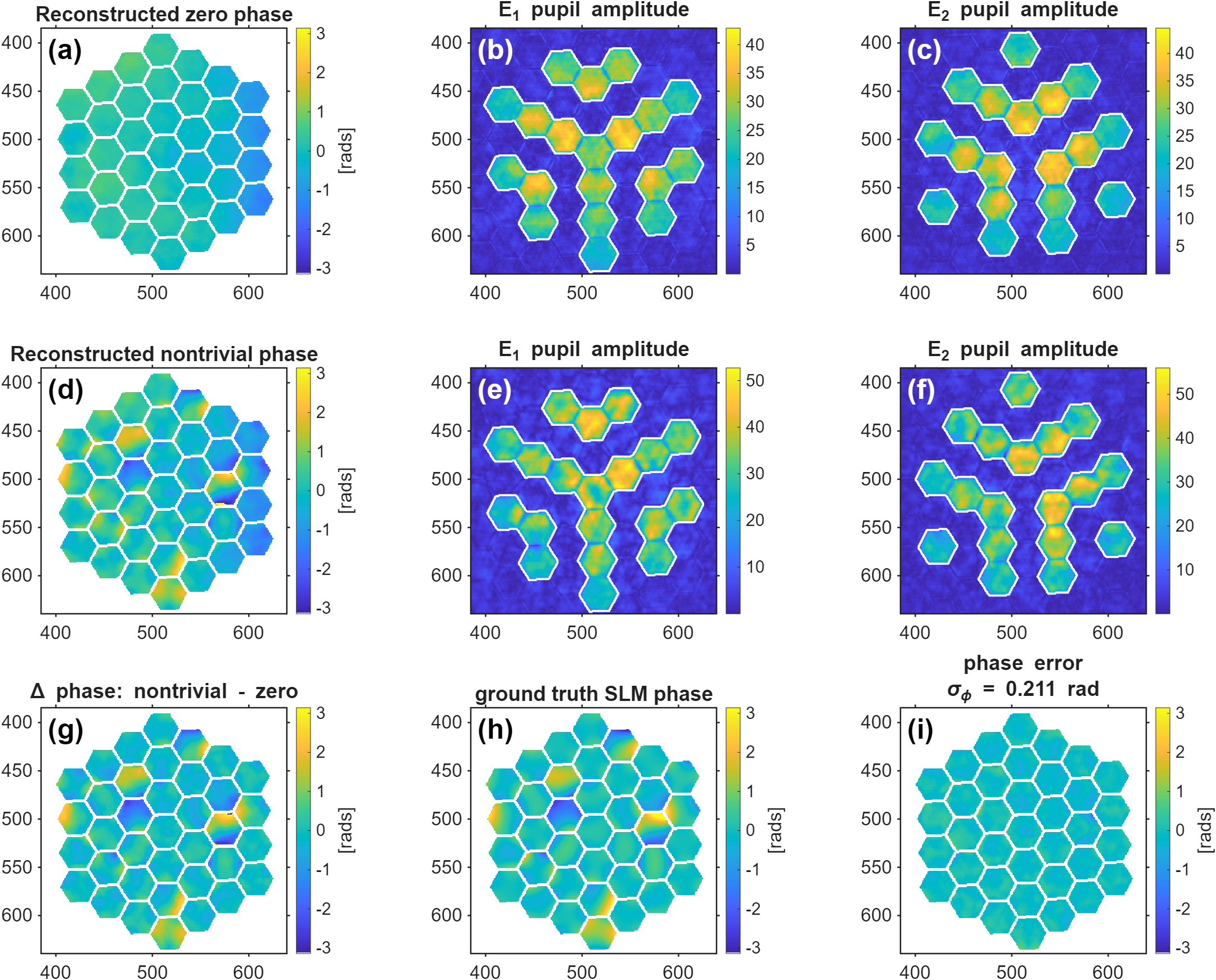}
    \vspace{1mm}
    \caption{\textbf{DBFH reconstruction results.} (a) The reconstructed pupil phase and (b-c) the odd- and even-segment field amplitudes, $E_1$ and $E_2$, for the reference zero SLM phase. (d-f) The nontrivial SLM phase and amplitude reconstruction. The bottom row compares the reconstructed differential phase (g), $\Delta\phi=\phi_2-\phi_1$, with the SLM phase pattern ground truth (h) and shows the residual phase error (i). The standard deviation of the residual over the full segmented pupil is $\sigma_{\phi}=0.211~\mathrm{rad}$.}
    \label{fig:DBFH results}
\end{figure}

Figure \ref{fig:DBFH results} shows the pupil plane-field as recovered by DBFH. In the top row the relatively flat reconstructed reference phase, along with the amplitude solutions. Note how the even and odd amplitudes solutions respect the CS constraints by leaking minimal energy to the complementary sub-aperture. The inter-segment separator vanes are also visible in this solution but are not enforced by the equations, demonstrating the ability of DBFH to recover the full complex field in the pupil plane. The middle row shows the phase and amplitude solutions of the applied non-trivial test phase. We compare the wrapped difference between the two phase solutions (bottom left) to the SLM ground truth phase (bottom center) to get the reconstruction phase error (bottom right). 

In this challenging case DBFH achieves RMSE of $0.21$rad. However, this error is a result of two reconstructions (reference and test), hence a single DBFH solution RMSE is $0.15$rad corresponding to $15$nm. This result is in close agreement with the RMSE expected of this SNR as published in our numerical study \cite{pedatzur2026double}. Importantly, the DBFH phase RMSE is expected to improve significantly with better SNR which cannot be achieved using this setup. 

\begin{figure}
    \centering
    \includegraphics[width=0.9\linewidth]{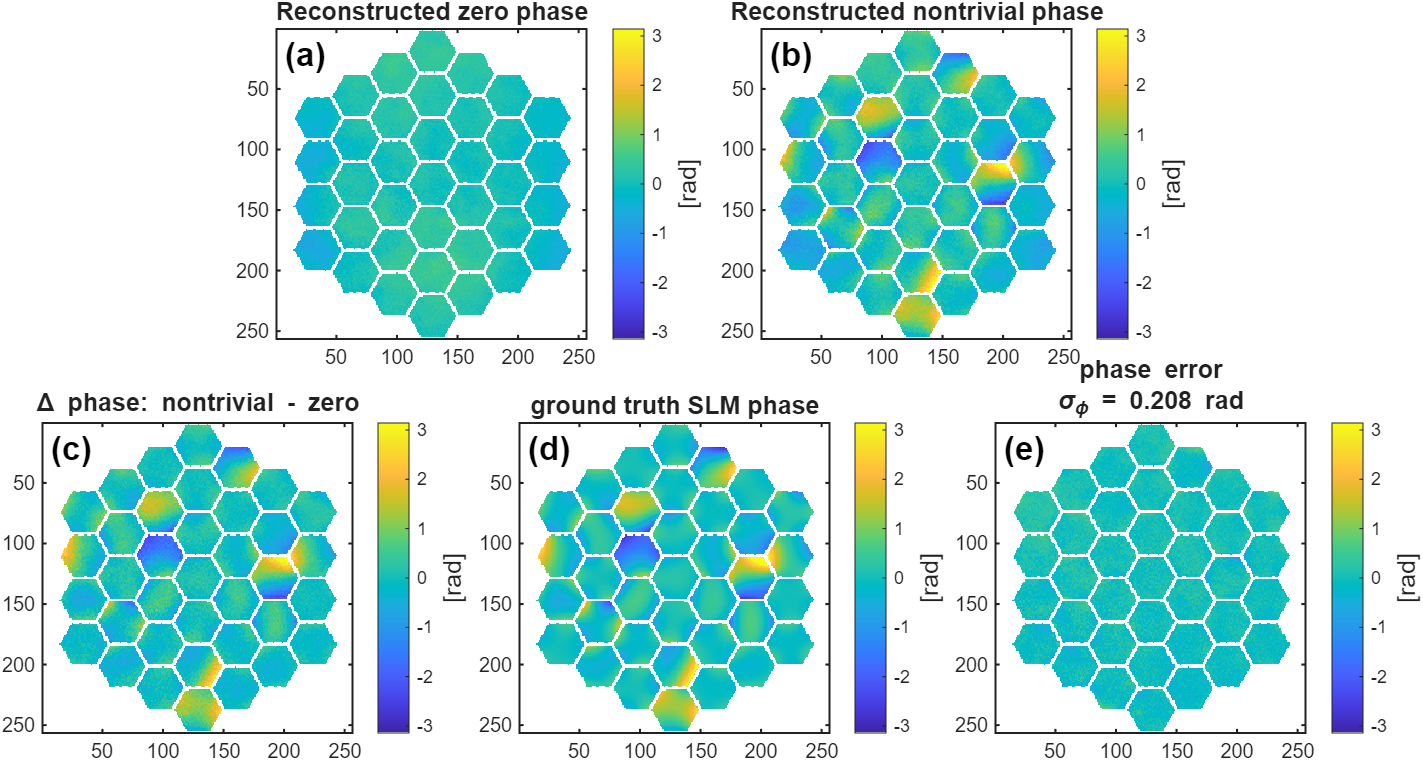}
    \vspace{1mm}
    \caption{\textbf{FDPR reconstruction results.} Focal diverse phase retrieval from $-$8, $-$4, 0, $+$4, and $+$8 waves defocus images. (a-b) FDPR reconstructions corresponding to the zero reference and nontrivial phase patterns. Bottom: differential phase (c) along with the ground truth SLM phase (d) and the residual phase error (e). The standard deviation of the residual over the full segmented pupil is $\sigma_{\phi}=0.208~\mathrm{rad}$.}
    \label{fig:PD results}
\end{figure}

Next, we apply FDPR using the exact same reference and non-trivial phase maps. The phase retrieval algorithm we employ is similar to the one used in the commissioning of the JWST \cite{acton2022phasing,dean2006phase}. Figure. \ref{fig:PD results} shows the defocused PSF inputs for both reference zero phase and the nontrivial WFE, and the corresponding optimizer solutions. As before, we subtract the reference phase solution from the non-trivial solution and compare to the SLM GT. Using FDPR we achieve RMSE slightly below $0.21$rad, similarly corresponding to a single phase reconstruction RMSE of $15$nm. Remarkably, both DBFH and FDPR achieve similar levels of precision, for this non-trivial phase pattern as well as other tested phases. This achievement situates DBFH in one line with well-established and trusted focal plane phase retrieval methods.

\begin{figure}
    \centering
    \includegraphics[width=0.8\linewidth]{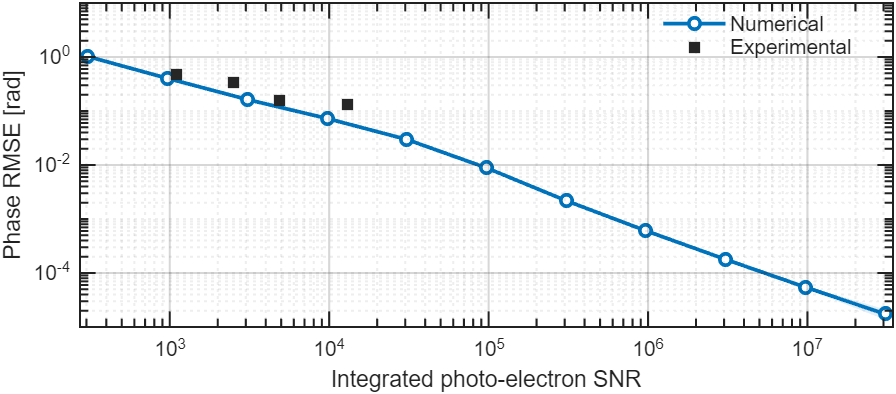}
    \vspace{1mm}
    \caption{\textbf{Phase solution RMSE for different SNR.}  DBFH phase solution RMSE for growing values of integrated photo-electron SNR, showing expected values from a numerical simulation including only Poissonian noise, and our experimental results for 1,10,100, and 1000 co-adds.}
    \label{fig: SNR sweep}
\end{figure}

Next, we sweep through various SNR conditions by applying DBFH to a series of intensity images from an increasing number of co-added frames, i.e. 1,10,100,1000. Based on previous numerical results \cite{pedatzur2026double}, we expect the asymptotic $1/\sqrt{N}$ improvement in RMSE to dominate only in significantly higher SNR values. Figure \ref{fig: SNR sweep} shows the expected RMSE vs. SNR curve from a numerical study including only Poissonian noise, along with our experimental SNR sweep points. Clearly, at higher frame co-adds a second noise source (e.g. read noise) or bias (e.g. model mismatch, source drift) dominates the error and prevents lower phase RMSE.

For a $1024\times1024$ reconstruction containing 1036957 complex unknowns, CPU execution on an Intel(R) Core(TM) i7-8700K CPU @ 3.70GHz required a median of 10.57 s for a complete DBFH reconstruction, with a 25th--75th percentile range of 10.54--10.65 s. This timing includes data preparation, construction of the linear system and preconditioner, LSQR solution, and pupil reconstruction. After the system and preconditioner had been assembled, repeated LSQR solves required a median of 10.28 s, with a 25th--75th percentile range of 10.24--10.34 s, and converged in a median of 39 iterations. Timings were obtained from 20 runs in MATLAB R2025b following warm-up runs.

Compared with FDPR, DBFH has several qualitative shortcomings that must be clarified. First, due to the finite full-well capacity, the in-focus inputs of DBFH require a much higher number of frames to achieve a similar count of photo-electrons, compared to FDPR's defocused images. This means that readout noise and dark-current which grow linearly with the frame count need to be suppressed by high full-well capacity and low-noise sensors (as can be expected for a flagship space telescope). The higher frame count may also extend the duration of acquisition session, perhaps conflicting with source stability. This was in fact a limiting case in our bench experiment, which prevented us from reaching higher number of useful co-adds. Second, as DBFH is essentially an interferometric method, its  algebra combines intensities from separate measurements. Exposure normalization, source brightness, detector gain, background subtraction, and pupil-mask throughput must therefore remain consistent.

DBFH offers several practical and computational advantages over FDPR. Its reconstruction is formulated as a linear inverse problem, enabling efficient and deterministic recovery without iterative nonlinear optimization. It directly solves for the complex pupil field, and does not assume a uniform pupil illumination or a low-dimensional aberration model. In applications where introducing controllable phase elements into the beam is undesirable, DBFH provides an \textit{in-situ} alternative based solely on amplitude masking.

\section{Binary Source} \label{section: binary source}

To better constrain WFS algorithms, it is a common practice to use a single bright and unresolved point source, such as an isolated reference bright star. This often requires a space telescope to point away from the science object of interest and to an ideal calibration scene. The appearance of multiple mutually incoherent PSFs in the FOV, or even incoherent background from dust will limit the ability of any interferometric method. Since DBFH consumes only in-focus images, the sensor area occupied by the PSF of a single source is significantly smaller compared to defocused images. While working with focused PSFs is disadvantageous in terms of finite full-well capacity and accumulation of read-out noise, it does allow accommodating multiple non-overlapping PSFs in the FOV. Here, we demonstrate DBFH wavefront sensing using an equal-brightness binary source.

\begin{figure}
    \centering
    \includegraphics[width=1\linewidth]{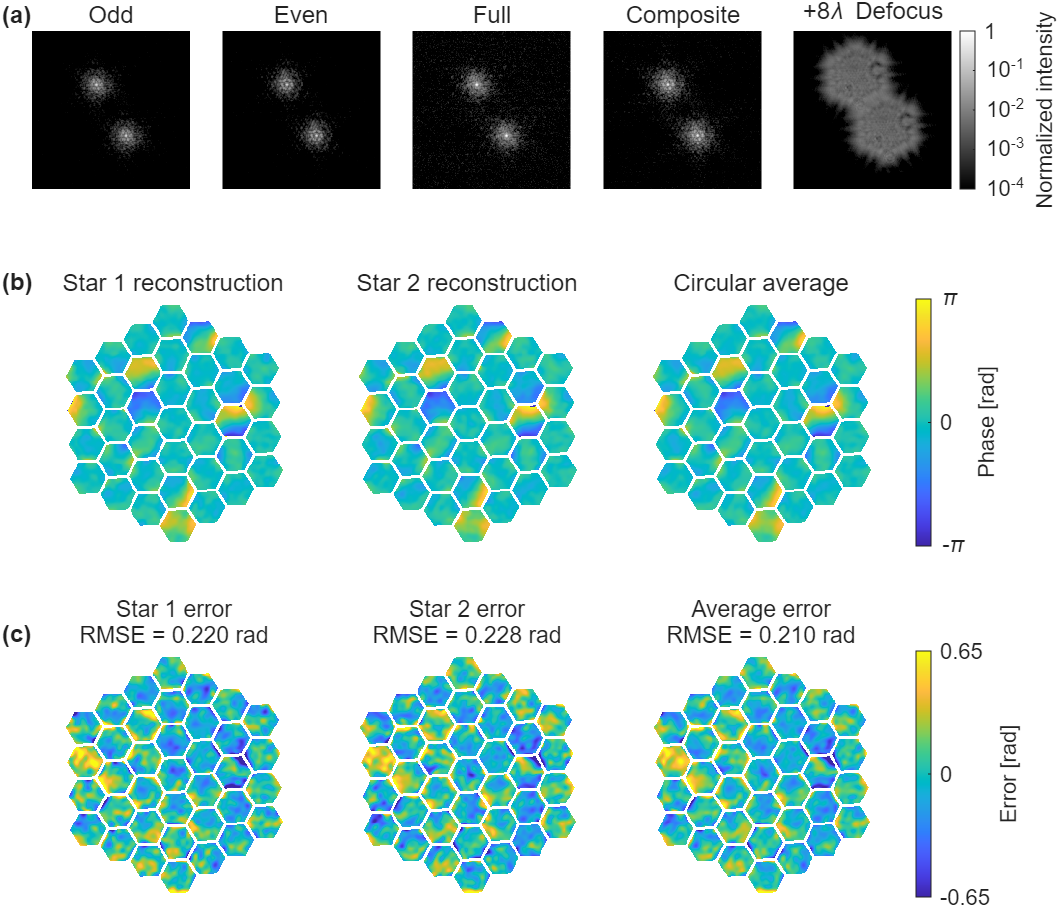}
    \vspace{1mm}
    \caption{\textbf{Binary Source Results.} (a) Focal-plane intensity measurements of a mutually incoherent binary source acquired using the odd, even, full, and in-quadrature DBFH pupil masks, alongside an $8\lambda$ defocused image representative of a FDPR measurement. The angular separation between the sources is $\sim50~ \lambda/D$, and all intensities are normalized and displayed using the same logarithmic scale. (b) Differential pupil-phase reconstructions obtained independently from each star and their equal-weight circular average. (c) Wrapped phase errors relative to the SLM ground truth (same as single-source).  Combining the two independent source reconstructions reduces the RMSE from 0.220 and 0.228 rad to 0.210 rad. This improvement is consistent with the numerical prediction for a twofold increase in detected intensity at this operating point, where the asymptotic $\sqrt{2}$ reduction in RMSE is not yet expected.}
    \label{fig:binary}
\end{figure}

Individual stars in a binary system are mutually incoherent, so their focal-plane intensities add directly. A binary scene can therefore be synthesized by shifting and summing independently acquired single-source images with matched exposure and photon statistics. This procedure correctly reproduces Poisson noise, since the sum of independent Poisson variables is itself Poisson-distributed. It does, however, overestimate the readout-noise contribution and is therefore justified only when Poisson noise dominates. Using offset copies of the same input images amounts to adding only redundant information (does not contribute new information about the wavefront) and is not expected to improve the phase RMSE. We chose to sum a 100 co-adds DBFH input images together with an offset and $\times \frac{1}{10}$ Poisson-thinned 1000 co-adds images. These two image sets are statistically independent observations of the same wavefront and therefore provide non-redundant information. Fig. \ref{fig:binary}(a) shows the fused DBFH inputs with separation of $\sim 50~\lambda/D$, along with a $+8\lambda$ defocus image (addition) emphasizing the advantage of in-focus analysis. The strong overlap between the defocused PSFs illustrates the spatial advantage of using in-focus measurements.

In this simple demonstration we simply apply DBFH to the two PSFs independently, followed by a cyclic averaging of the estimated wavefronts. Alternative approaches include registering and summing the centered PSFs before applying DBFH, or constructing a joint linear system containing measurements from both sources. These latter two methods require accurate knowledge of the relative source displacement, whereas independently reconstructed wavefronts can be detrended before averaging.

Fig. \ref{fig:binary}(b) shows the individual wavefront reconstructions, following a procedure similar to the one applied for a single source (section \ref{section: results}), together with their circular average. Fig. \ref{fig:binary}(c) reports the corresponding RMSE differences relative to the SLM ground-truth, showing the expected and modest improvement to the RMSE in the joint case, as expected from fig. \ref{fig: SNR sweep}. More importantly, this result demonstrates that DBFH wavefront sensing remains feasible in a binary field. The independent companion measurement provides a secondary benefit by reducing the reconstruction RMSE, much like additional captures or a longer effective integration time.

\section{Conclusions}\label{section: conclusions}

In this work, we experimentally validated DBFH as a focal-plane wavefront-sensing method for future segmented space telescopes, with particular relevance to the Habitable Worlds Observatory. Using a programmable SLM, we applied and reconstructed nontrivial segmented-aperture wavefronts and compared the DBFH estimates with both the known SLM phase and FDPR reconstructions, a widely used benchmark for focal-plane wavefront sensing. For the test phases considered, DBFH achieved reconstruction precision comparable to FDPR.

We further evaluated DBFH over a range of exposure levels and found that the measured phase RMSE followed the numerical prediction in the Poisson-noise-dominated regime. Compared with FDPR on equivalent inputs, DBFH achieves comparable WFE precision. This comparison highlighted the distinct trade-offs between the two methods, including in-focus versus defocused measurements and amplitude masking versus phase modulation. Because DBFH is formulated as an algebraically linear inverse problem, it offers computational advantages over iterative phase-retrieval approaches. Finally, we demonstrated that the smaller area occupied by the in-focus PSF on the detector allows spatially separated stars within the same field of view to be processed independently. The additional statistically independent measurement from a companion source improves the reconstruction precision. Together, these results establish an experimental proof of concept for DBFH and support its advancement to TRL 3 as a wavefront-sensing method for segmented space telescopes.

\section{Disclosures}\label{section: disclosures}

The authors declare no conflicts of interest.



\bibliography{experimental_validation_bib.bib}

@inproceedings{acton2022phasing,
  title={Phasing the Webb telescope},
  author={Acton, D Scott and Knight, Scott and Carrasquilla, Maria and Weiser, Nick and Masciarelli, Michaela and Jurczyk, Sarah and Rapp, Greg and Mueckay, Julio and Wolf, Erin and Murphy, Jess and others},
  booktitle={Space Telescopes and Instrumentation 2022: Optical, Infrared, and Millimeter Wave},
  volume={12180},
  pages={329--352},
  year={2022},
  organization={SPIE}
}

@article{gonsalves1982phase,
  title={Phase retrieval and diversity in adaptive optics},
  author={Gonsalves, Robert A},
  journal={Optical Engineering},
  volume={21},
  number={5},
  pages={829--832},
  year={1982},
  publisher={SPIE}
}

@inproceedings{dean2006phase,
  title={Phase retrieval algorithm for JWST flight and testbed telescope},
  author={Dean, Bruce H and Aronstein, David L and Smith, J Scott and Shiri, Ron and Acton, D Scott},
  booktitle={Space telescopes and instrumentation I: optical, infrared, and millimeter},
  volume={6265},
  pages={314--330},
  year={2006},
  organization={SPIE}
}

@inproceedings{gonsalves2018phase,
  title={Phase diversity: math, methods and prospects, including sequential diversity imaging},
  author={Gonsalves, Robert A},
  booktitle={Unconventional Optical Imaging},
  volume={10677},
  pages={335--345},
  year={2018},
  organization={SPIE}
}

@inproceedings{acton2004james,
  title={James Webb Space Telescope wavefront sensing and control algorithms},
  author={Acton, Daniel Scott and Atcheson, Paul D and Cermak, Mick and Kingsbury, Lana K and Shi, Fang and Redding, David C},
  booktitle={Optical, Infrared, and Millimeter Space Telescopes},
  volume={5487},
  pages={887--896},
  year={2004},
  organization={SPIE}
}

@article{pedatzur2019double,
  title={Double-blind holography of attosecond pulses},
  author={Pedatzur, O and Trabattoni, A and Leshem, B and Shalmoni, H and Castrovilli, MC and Galli, M and Lucchini, M and M{\aa}nsson, E and Frassetto, F and Poletto, L and others},
  journal={Nature Photonics},
  volume={13},
  number={2},
  pages={91--95},
  year={2019},
  publisher={Nature Publishing Group UK London}
}

@article{leshem2016direct,
  title={Direct single-shot phase retrieval from the diffraction pattern of separated objects},
  author={Leshem, Ben and Xu, Rui and Dallal, Yehonatan and Miao, Jianwei and Nadler, Boaz and Oron, Dan and Dudovich, Nirit and Raz, Oren},
  journal={Nature communications},
  volume={7},
  number={1},
  pages={10820},
  year={2016},
  publisher={Nature Publishing Group UK London}
}

@article{raz2014direct,
  title={Direct phase retrieval in double blind Fourier holography},
  author={Raz, Oren and Leshem, Ben and Miao, Jianwei and Nadler, Boaz and Oron, Dan and Dudovich, Nirit},
  journal={Optics express},
  volume={22},
  number={21},
  pages={24935--24950},
  year={2014},
  publisher={Optica Publishing Group}
}

@article{raz2011vectorial,
  title={Vectorial phase retrieval for linear characterization of attosecond pulses},
  author={Raz, Oren and Schwartz, Osip and Austin, Dane and Wyatt, AS and Schiavi, Andrea and Smirnova, Olga and Nadler, Boaz and Walmsley, Ian A and Oron, Dan and Dudovich, Nirit},
  journal={Physical review letters},
  volume={107},
  number={13},
  pages={133902},
  year={2011},
  publisher={APS}
}

@article{raz2013vectorial,
  title={Vectorial phase retrieval of 1-D signals},
  author={Raz, Oren and Dudovich, Nirit and Nadler, Boaz},
  journal={IEEE Transactions on Signal Processing},
  volume={61},
  number={7},
  pages={1632--1643},
  year={2013},
  publisher={IEEE}
}

@article{pedatzur2026double,
  title={Double blind Fourier holography for high-precision wavefront sensing in segmented mirror telescopes},
  author={Pedatzur, Oren},
  journal={Optics Express},
  volume={34},
  number={5},
  pages={8223--8234},
  year={2026},
  publisher={Optica Publishing Group}
}

@article{pedatzur2026differential,
  title={Differential Double Blind Fourier Holography (diff-DBFH)},
  author={Pedatzur, Oren and Chuss, David T},
  journal={arXiv preprint arXiv:2609.30183},
  year={2026}
}

@book{NASEM2021Astro2020,
  title        = {Pathways to Discovery in Astronomy and Astrophysics for the 2020s},
  author       = {{National Academies of Sciences, Engineering, and Medicine}},
  year         = {2021},
  publisher    = {The National Academies Press},
  address      = {Washington, DC}
}

@article{juanola2022modeling,
  title={Modeling and performance analysis of the LUVOIR coronagraph instrument},
  author={Juanola-Parramon, Roser and Zimmerman, Neil T and Pueyo, Laurent and Bolcar, Matthew and Gong, Qian and Groff, Tyler and Krist, John and Roberge, Aki and Ruane, Garreth and Stark, Christopher},
  journal={Journal of Astronomical Telescopes, Instruments, and Systems},
  volume={8},
  number={3},
  pages={034001--034001},
  year={2022},
  publisher={Society of Photo-Optical Instrumentation Engineers}
}

@inproceedings{juanola2019evaluating,
  title={Evaluating the LUVOIR coronagraph sensitivity to telescope aberrations},
  author={Juanola-Parramon, Roser and Zimmerman, Neil and Groff, Tyler and Pueyo, Laurent and Rizzo, Maxime and Bolcar, Matthew},
  booktitle={2019 IEEE Aerospace Conference},
  pages={1--8},
  year={2019},
  organization={IEEE}
}

\end{document}